# Appearance of ferroelectricity in thin films of incipient ferroelectric.

Eugene A. Eliseev, Maya D.Glinchuk, and Anna N.Morozovska[*],

Institute for Problems of Materials Science, NAS of Ukraine,
Krjijanovskogo 3, 03142 Kiev, Ukraine, eliseev@i.com.ua, glin@materials.kiev.ua

The consideration of size-induced ferroelectric-paraelectric phase transition for conventional and incipient ferroelectrics thin films with perovskite structure was carried out in phenomenological theory framework of Ginsburg-Landau-Devonshire. The more general form of surface free energy expansion that includes intrinsic surface stress tensor, surface piezoelectric effect and electrostriction as well as quadratic and quartic powers of surface polarization has been considered. The analytical expressions for thickness dependence of transition temperature was derived both for the conventional and incipient ferroelectrics. It was shown that although there is no ferroelectricity in the bulk incipient ferroelectrics it appears in thin film for the negative extrapolation length that is realized e.g. at positive surface stress coefficient and negative or zero misfit strain. In our consideration we came to the conclusion about thickness induced ferroelectricity in incipient ferroelectrics $KTaO_3$ at room temperature for the thin enough films. The similar surface effects can influence strongly on the phase transitions in the thin films of another incipient ferroelectrics, $SrTiO_3$.

PACS: 77.80.-e, 77.84.Dy, 68.03.Cd, 68.35.Gy

## 1. INTRODUCTION

In the last years much attention was paid to the consideration of ferroelectric thin films and their multilayers [1]. Besides the deposition conditions the main topics were related to the influence of film thickness, substrate, electrodes and ambient conditions on the film properties [2]. Practically all the properties of the films were shown to be different from those in the bulk because of the films structure and surface influence. These facts as well as the possibility to manage the film properties by changing aforementioned factors opens the way for obtaining the films with properties useful for modern applications such as dynamic random access memories (DRAM) and non-volatile ferroelectric random access memories (NVFRAM). The progress in this direction depends strongly on the understanding of physical mechanisms, which are responsible for the properties peculiarities.

---

[*] Corresponding author: morozo@i.com.ua, permanent address: V. Lashkarev Institute of Semiconductor Physics, NAS of Ukraine, 41, pr. Nauki, 03028 Kiev, Ukraine

In the most of the theoretical papers based on phenomenological theory polarization gradient, depolarization field energy as well as elastic energy and electrostriction contribution is included into the free energy functional of ferroelectric thin film, while the surface energy usually contains only powers of polarization (up to forth one). Rather often surface energy and gradient energy contribution are neglected. Being rigorous this approach allows one to consider the properties of the thick films of multiaxial ferroelectrics on substrate with respect to misfit strain (see e.g. [3], [4]), however one can hardly apply these results to the thin films. Recently we included mismatch strain as well as surface piezoelectric effect into the surface energy and shown that they can lead to the appearance of the built-in field and the phase transition smearing in the thin ferroelectric films [5].

The surface energy and the surface intrinsic stress effects on spontaneous formation of periodically ordered nanometer-scale structures at crystal surfaces is well understood [6]. It is obvious that this approach can be extended to the thin ferroelectric films, allowing for the polarization coupling with elastic stress, since characteristic feature of the nanoscale structures phenomenological description is the surface energy contribution that becomes comparable with the bulk one and can exceed it under size decrease.

This consideration should be based on the elastic problems solutions for the freestanding films and for the films on the substrate allowing for the correct the mechanical boundary conditions. To the best of our knowledge there was no detailed and general consideration of these problems for ferroelectric films up to now. For the incipient ferroelectrics, which are known to be in paraelectric phase up to zero Kelvin, the main question is the possibility of ferroelectric phase appearance in the films of incipient ferroelectrics.

The phenomenological model for epitaxial films of incipient ferroelectric $SrTiO_3$ was considered earlier for homogeneously strained film taking into account three components of polarization and structural order parameter, but with size, surface and depolarization field being ignored [4]. To our mind these simplifications may explain why the predicted by Pertsev et al. [4] temperature of transition to ferroelectric phase is appeared to be significantly lower than that determined experimentally [7]. Recently Li et al. [8] used the above-mentioned model for the description of domain structure in thin films of $SrTiO_3$ with respect to the gradient energy associated with domain walls. They showed that variations in the reported values of bulk free energy expansion coefficients lead to the wide range of transition temperature values.

In this paper we will consider thin films of both the conventional and incipient ferroelectric of perovskite structure in the framework of Ginsburg-Landau-Devonshire phenomenological approach with respect to the surface energy, polarization gradient (correlation) energy, depolarization field energy as well as elastic energy and electrostriction contribution included in the free energy functional. Into surface energy we include the polarization even powers as well as surface piezoelectric effect contribution, electrostriction and intrinsic surface stress. The latter allows us to demonstrate that under the favorable conditions surface stress essentially increases the transition temperature and may induce ordered phase in incipient ferroelectrics.

## 2. FREE ENERGY FUNCTIONAL AND MECHANICAL BOUNDARY CONDITIONS

For perovskite symmetry Hibbs bulk free energy expansion on polarization $\mathbf{P} = (0, 0, P_3)$ and stress tensor $\hat{\sigma}$ components powers has the form: [9]

$$G_V = \int_V d^3r \left( \begin{array}{l} \dfrac{a_1(T)}{2}P_3^2 + \dfrac{a_{11}}{4}P_3^4 + \dfrac{a_{111}}{6}P_3^6 + \dfrac{g}{2}(\nabla P_3)^2 \\ - P_3\left(E_0 + \dfrac{E_3^d}{2}\right) - Q_{ij33}\sigma_{ij}P_3^2 - \dfrac{1}{2}s_{ijkl}\sigma_{ij}\sigma_{kl} \end{array} \right) \tag{1a}$$

In accordance with Landau-Ginzburg phenomenological theory only coefficient $a_1(T)$ before quadratic term of order parameters depends on temperature. High order coefficients $a_{11}$ and $a_{111}$ are supposed temperature independent, $g$ determines correlation energy. Since the polarization of bulk system is finite and homogeneous, one should also suppose that $g > 0$ and $a_{111} > 0$. Sign of $a_{11}$ depends on the phase transition order, namely $a_{11} < 0$ for the first order and $a_{11} > 0$ for the second order phase transition. Here $E_3^d$ is depolarization field [10], $E_0$ is external field z-components; $Q_{ijkl}$ and $s_{ijkl}$ are the components of electrostriction and elastic compliances tensor correspondingly.

The surface free energy expansion on $P_3$ and $\hat{\sigma}$ components has the form

$$G_S = \sum_{i=1}^{2} \int_i d^2r \left( \begin{array}{l} \dfrac{a_1^{(i)}}{2}P_3^2 + \dfrac{a_{11}^{(i)}}{4}P_3^4 - \tau_{jk}^{(i)}\sigma_{jk} - \dfrac{v_{jklm}^{(i)}}{2}\sigma_{jk}\sigma_{lm} + \\ - d_{3jk}^{(i)}\sigma_{jk}P_3 - q_{jk33}^{(i)}\sigma_{jk}P_3^2 \end{array} \right) \tag{1b}$$

Here $i = 1, 2$ denote the number of the surface; surface displacement tensor $\tau_{jk}^{(i)}$ is related with the intrinsic surface stress tensor $\mu_{\alpha\beta}^{(i)}$ [6], [11] via the elastic compliances tensor $s_{ijkl}$ in accordance with Hooke law as $\tau_{jk}^{(i)} = \mu_{\alpha\beta}^{(i)}s_{\alpha\beta jk}$; $v_{jklm}^{(i)}$ is the surface elastic compliances. Greek characters label two-dimensional indices in the surface plane, whereas Roman indices are three-dimensional ones. $u_{jk} = (\partial u_j/\partial x_k + \partial u_k/\partial x_j)/2$ is the strain tensor and $u_j$ is displacement vector components.

For the sake of simplicity hereinafter we consider the case of isotropic solid, where the symmetry of surface stress tensors are isotropic, namely $\tau_{jk}^{(i)} = \tau^{(i)}\delta_{jk}$, $\mu_{jk}^{(i)} = \mu^{(i)}\delta_{jk}$ and $v_{jklm}^{(i)} = v_{12}^{(i)}\delta_{jk}\delta_{lm} + \left(v_{11}^{(i)} - v_{12}^{(i)}\right)\left(\delta_{jl}\delta_{km} + \delta_{jm}\delta_{kl}\right)/2$ ($\delta_{jk}$ is the Kroneker symbol). After summation this leads to $\tau_{jk}^{(i)}\sigma_{jk} = \tau^{(i)}(\sigma_{11} + \sigma_{22} + \sigma_{33})$ and $v_{jklm}^{(i)}\sigma_{jk}\sigma_{lm} = v_{11}^{(i)}(\sigma_{jj})^2 + \left(v_{11}^{(i)} - v_{12}^{(i)}\right)\sigma_{jk}\sigma_{kj}$.

It should be noted that physical quantities $\tau^{(i)}$ and $v^{(i)}_{11,12}$ can be either positive or negative. The sign of surface stress tensor components $\mu^{(i)}_{\alpha\beta}$ depends on the chemical properties of the ambient material, the presence of oxide or interface layer. Taking into account, that there exist surface layers/interfaces with chemical, structural and polar properties different from those of the bulk, hereinafter we consider both positive and negative values of $\mu^{(i)}_{\alpha\alpha}$.

Tensor $d^{(i)}_{3jk}$ is related to surface piezoelectric effect existing even in a cubic symmetry lattice near the film surface (since inversion center disappears in surface normal direction, see e.g. [12], [5], [13]), $q^{(i)}_{jklm}$ is the surface electrostriction tensor, which symmetry is the same as bulk electrostriction tensor $Q_{jklm}$ has, but their signs and absolute values can be different.

Free energy (1) is minimal when polarization $P_3$ and relevant stress tensor components $\sigma_{jk}$ are defined at the nanostructure boundaries [14]. Under such conditions, one should solve equation of state $\partial G/\partial \sigma_{jk} = -u_{jk}$ that determines the generalized Hooke's law for the considered system.

For the cases of the clamped system with defined displacement components (or with mixed boundary conditions) one should find the equilibrium state as the minimum of the Helmholtz free energy $F_V + F_S$ ($F_V = G_V + \int_V d^3r \cdot u_{jk}\sigma_{jk}$ and $F_S = G_S + \int_S d^2r \cdot u_\alpha \sigma_{\alpha k} n_k$) originated from Legendre transformation of $G$ [11], [3].

Equilibrium equations of state could be obtained after variation of the Helmholtz energy on displacement $u_j$, Gibbs energy on stress $\sigma_{ij}$, polarization $P_3$ and its derivatives:

$$\frac{\partial \sigma_{ij}}{\partial x_i} = 0, \quad Q_{ij33}P_3^2 + s_{ijkl}\sigma_{kl} = u_{ij}, \quad (a_1 - Q_{ij33}\sigma_{ij})P_3 + a_{11}P_3^3 + a_{111}P_3^5 - g\frac{\partial^2 P_3}{\partial x_k \partial x_k} = E_0 + E_3^d. \quad (2)$$

Eqs.(2) should be supplemented by the mechanical boundary conditions for strain (or stress) on the nanostructure surfaces.

System consisting of thin epitaxial films on the thick or rigid substrate represents the mixed type of mechanical boundary conditions, since the top surface is *free* (i.e. the normal stress components are zero) and the film substrate interface is *clamped* (i.e. the strain components $u_{\alpha\beta}$ should be determined) [3]. For particular case when the seeding misfit strain $u^m_{\alpha\beta}$ intergrown through the thin film, one obtains that $u_{\alpha\beta} = u^m_{\alpha\beta}$ on both surfaces.

*2.1. Elastic problem for thin unipolar perovskite film on a rigid substrate*

We assume that at the film-substrate surface ($z = -l/2$) the strain is induced by film-substrate lattice mismatch as following:

$$u_{11} = u_{22} = u_m, \quad u_{12} = 0. \quad (3a)$$

At the film-ambient free surface ($z = +l/2$) the stress is determined as

$$\sigma_{33} = \sigma_{31} = \sigma_{32} = 0. \tag{3b}$$

In the considered case of a single-domain perovskite thin film with polarization $\mathbf{P} = (0,0,P_3)$ on a rigid cubic substrate, internal elastic fields are homogeneous so that the above conditions hold throughout the film volume for the film with thickness $l$ smaller than the critical thickness of misfit dislocations appearance $l_d$. Otherwise, $u_m$ should the substituted by $u_m^* = u_m l_d / l$ in accordance with the Speck and Pompe model [15]. Thus, the solution of the "mixed" elastic problem, obtained after minimization of $\dfrac{\partial G_V}{\partial \sigma_{jk}} = -u_{jk}$ has the form:

$$u_{11} = u_{22} = u_m, \quad u_{33} = 2s_{12}\frac{u_m - Q_{12}P_3^2}{s_{11} + s_{12}} + Q_{11}P_3^2, \quad u_{23} = u_{13} = u_{12} = 0, \tag{4a}$$

$$\sigma_{11} = \sigma_{22} = \frac{u_m - Q_{12}P_3^2}{s_{11} + s_{12}}, \quad \sigma_{12} = \sigma_{31} = \sigma_{32} = \sigma_{33} = 0. \tag{4b}$$

Where the perovskite symmetry ($Q_{1133} = Q_{2233} = Q_{12}$, $Q_{3333} = Q_{33} = Q_{11}$) and Voigt notation (xx=1, yy=2, zz=3, zy=4, zx=5, xy=6) are used.

*2.2. Freestanding unipolar perovskite film*

In particular case of a *freestanding* thin film one should use the boundary conditions $\sigma_{3k} = 0$ on both free surfaces $z = \pm l/2$. It is appeared that the solution of the corresponding elastic problem

$$\sigma_{jk} = 0, \quad u_{11} = u_{22} = Q_{12}P_3^2, \quad u_{33} = Q_{11}P_3^2, \quad u_{23} = u_{13} = u_{12} = 0. \tag{5}$$

Note, that solution (5) is the spontaneous deformation of bulk material and thus it does not transforms into the solution (4) at $u_m \to 0$, since they correspond to the different boundary conditions at $z = -l/2$, namely zero strain for clamped film or zero stress for the free one.

## 3. EULER-LAGRANGE EQUATION FOR A FREESTANDING THIN FILM

The behavior of a freestanding film can be described in the following way. Substituting the solution (5) of the mechanical problem into the Gibbs free energy (1) we obtained:

$$\frac{G}{S} = \begin{pmatrix} \int\limits_{-l/2}^{l/2} dz \left( \dfrac{a_1}{2}P_3^2 + \dfrac{a_{11}}{4}P_3^4 + \dfrac{a_{111}}{6}P_3^6 + \dfrac{g}{2}\left(\dfrac{dP_3}{dz}\right)^2 - P_3\left(E_0 + \dfrac{E_3^d}{2}\right) \right) \\ + \dfrac{a_1^{(1)}}{2}P_3^2\left(-\dfrac{l}{2}\right) + \dfrac{a_{11}^{(1)}}{4}P_3^4\left(-\dfrac{l}{2}\right) + \dfrac{a_1^{(2)}}{2}P_3^2\left(\dfrac{l}{2}\right) + \dfrac{a_{11}^{(2)}}{4}P_3^4\left(\dfrac{l}{2}\right) \end{pmatrix} \tag{6}$$

The variation of free energy (6) leads to the Euler-Lagrange equation for polarization $P_3(z)$:

$$\begin{cases} a_1 P_3 + a_{11} P_3^3 + a_{111} P_3^5 - g\dfrac{d^2 P_3}{dz^2} = E_0 + E_3^d, \\ \left(P_3 + \lambda_1\left(\dfrac{dP_3}{dz} + \dfrac{a_{11}^{(1)}}{g} P_3^3\right)\right)\bigg|_{z=l/2} = 0, \quad \left(P_3 - \lambda_2\left(\dfrac{dP_3}{dz} - \dfrac{a_{11}^{(2)}}{g} P_3^3\right)\right)\bigg|_{z=-l/2} = 0. \end{cases} \quad (7)$$

where the extrapolation lengths $\lambda_i = g/a_1^{(i)}$ are introduced. For conventional ferroelectrics we supposed that $a_1(T) = \alpha_T(T - T_C^b)$, where $T_C^b$ is the bulk material transition temperature. The solution of linearized Eq. (7) corresponds to paraelectric phase, where polarization $P_3$ is proportional to the external field $E_0$. Then one can find the paraelectric phase dielectric permittivity as $\varepsilon_{33} = 4\pi dP_3/dE_0$. The condition of inverse permittivity tending to zero ($\varepsilon_{33}^{-1} \to 0$) determines the paraelectric phase instability point, which is the transition point for the phase transition of second order. Under the condition $\lambda_1 = \lambda_2 = \lambda$ exact expression for the ferroelectric transition temperature $T_f(l)$ is given by:

$$T_f(l,\lambda) = T_C^b - \dfrac{g}{\alpha_T l} \dfrac{2\sinh\left(l\sqrt{\pi/g}\right)}{\sqrt{g/4\pi}\cosh\left(l\sqrt{\pi/g}\right) + \lambda \sinh\left(l\sqrt{\pi/g}\right)}. \quad (8)$$

Despite technological difficulties, recently freestanding films of BaTiO$_3$ have been produced by Saad et al. [16]. They observed sharp maximum on the temperature dependence of the dielectric permittivity in contrast to the smeared ones typical for the films clamped on the substrate. The theoretical consideration [5] explains this result as the consequence of the internal electric field, proportional to the misfit strain. It is obvious that this field is absent for the freestanding film.

## 4. EULER-LAGRANGE EQUATION FOR THIN FILM ON A RIGID SUBSTRATE

Let us consider ferroelectric unipolar (i.e. $\mathbf{P} = (0,0,P_3)$) thin perovskite film with the thickness $l$ ($-l/2 \le z \le l/2$) on the thick substrate in the external electric field $\mathbf{E} = (0,0,E_0)$.

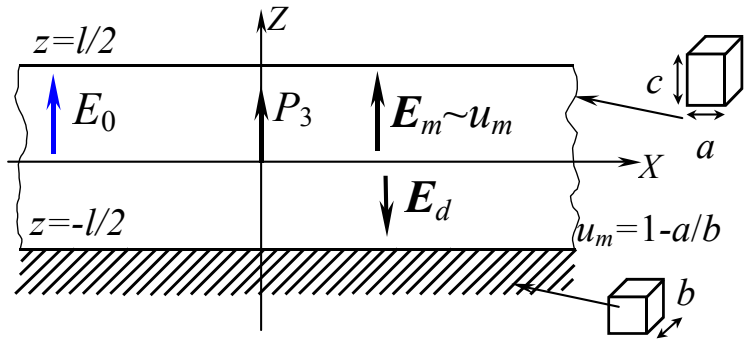

**FIG. 1.** Geometry of the film.

Substitution of solution (4) into the free energy (1) leads to the expression for its polarization-dependent part (see Appendix A):

$$F_V = S \int_{-l/2}^{l/2} dz \left( \left( \frac{a_1}{2} - \frac{2Q_{12}u_m}{s_{11}+s_{12}} \right) P_3^2 + \left( \frac{a_{11}}{4} + \frac{Q_{12}^2}{s_{11}+s_{12}} \right) P_3^4 + \frac{a_{111}}{6} P_3^6 + \frac{g}{2}\left(\frac{dP_3}{dz}\right)^2 - P_3\left(E_0 + \frac{E_3^d}{2}\right) \right) \quad (9a)$$

$$F_S = \sum_{i=1}^{2} \int_{S_i} dxdy \left( \left( \frac{a_1^{(i)}}{2} + 2\frac{\tau^{(i)}Q_{12} - q_{12}^{(i)}u_m}{s_{11}+s_{12}} - \frac{2(v_{11}^{(i)}+v_{12}^{(i)})}{(s_{11}+s_{12})^2} u_m Q_{12} \right) P_3^2 + \left( \frac{a_{11}^{(i)}}{4} - \frac{(v_{11}^{(i)}+v_{12}^{(i)})Q_{12}^2}{(s_{11}+s_{12})^2} + \frac{2q_{12}^{(i)}Q_{12}}{s_{11}+s_{12}} \right) P_3^4 - \frac{2d_{31}^{(i)}u_m P_3}{s_{11}+s_{12}} + \frac{2d_{31}^{(i)}u_m Q_{12} P_3^3}{s_{11}+s_{12}} \right) \quad (9b)$$

Here $d_{311}^{(i)} = d_{322}^{(i)} = d_{31}^{(i)}$ for perovskites.

Allowing for expression for depolarization field $E_3^d = 4\pi(\overline{P_3} - P_3)$ proposed by Kretschmer and Binder [17] (where $\overline{P}$ stands for the spatial average on the sample volume), variation of free energy (9) leads to the Euler-Lagrange equation for polarization $P_3(z)$:

$$\begin{cases} \left( a_1 - \frac{4Q_{12}u_m}{s_{11}+s_{12}} \right) P_3 + \left( a_{11} + \frac{Q_{12}^2}{s_{11}+s_{12}} \right) P_3^3 + a_{111} P_3^5 - g\frac{d^2 P_3}{dz^2} = E_0 + 4\pi(\overline{P_3} - P_3), \\ \left. \left( P_3 + \lambda_S^{(i)}\left( \frac{dP_3}{dz} + b_1 P_3^3 + c_1 P_3^2 \right) \right) \right|_{z=l/2} = -P_{m1}, \\ \left. \left( P_3 - \lambda_S^{(i)}\left( \frac{dP_3}{dz} - b_2 P_3^3 - c_2 P_3^2 \right) \right) \right|_{z=-l/2} = -P_{m2}. \end{cases} \quad (10)$$

Note, that Euler-Lagrange equation for freestanding film (7) could not be obtained as the limit of Eq. (10) at $u_m \to 0$ because of the differences in the solution of corresponding elastic problems (4) and (5).

In Eqs. (10) the boundary conditions have been rewritten via renormalized length $\lambda_S$ and surface polarization $P_{mi}$, namely:

$$\frac{1}{\lambda_S^{(i)}} = \frac{a_1^{(i)}}{g} + \frac{2}{g}\left( 2\frac{\tau^{(i)}Q_{12} - q_{12}^{(i)}u_m}{s_{11}+s_{12}} - \frac{2(v_{11}^{(i)}+v_{12}^{(i)})}{(s_{11}+s_{12})^2} u_m Q_{12} \right), \quad (11)$$

$$P_{mi} = \frac{2d_{31}^{(i)}u_m}{s_{11}+s_{12}} \frac{\lambda_S^{(i)}}{g}, \quad (12)$$

$$b_i = \frac{1}{g}\left( a_{11}^{(i)} - 4\frac{(v_{11}^{(i)}+v_{12}^{(i)})Q_{12}^2}{(s_{11}+s_{12})^2} + \frac{8q_{12}^{(i)}Q_{12}}{s_{11}+s_{12}} \right), \quad c_i = \frac{12d_{31}^{(i)}u_m Q_{12} P_3^3}{s_{11}+s_{12}}. \quad (13)$$

It should be noted that characteristic length $\lambda_S^{(i)}$ has the meaning of extrapolation length when surface piezoelectric effect is absent (i.e. $P_{mi} = 0$). Putting hereinafter $\tau^{(i)} = \mu^{(i)}(s_{11} + s_{12})$ and assuming that $q_{12}^{(i)} \approx Q_{12}h$ and $v_{11}^{(i)} + v_{12}^{(i)} \approx h(s_{11} + s_{12})$, where $h$ is the thickness of the surface layer (about several lattice constants), one can rewrite Eq. (11) in more simple form:

$$\frac{1}{\lambda_S^{(i)}} \approx \frac{1}{\lambda_i} + \frac{4}{g}Q_{12}\left(\mu^{(i)} - \frac{2u_m}{s_{11} + s_{12}}h\right) \quad (14)$$

Here $\lambda_i = g/a_1^{(i)}$ has the meaning of extrapolation length for the free standing film. Hereinafter we assume that the film surfaces have the same properties and omit superscript ($i$) for clearance. Using typical values of parameters $(s_{11} + s_{12}) \sim 10^{-12}$ 1/Pa, $\mu^S \sim (-50...+50)$Pa·m [18], $|u_m| < 10^{-2}$, $h$=0.4-0.8 nm, $Q_{12} \sim -10^{-1} \div 10^{-2}$ m$^4$/C$^2$, $g \sim 10^{-10}$ V m$^3$/C (see e.g. [19]), we obtained that the expression $\frac{4}{g}Q_{12}\left(\mu - \frac{2u_m}{s_{11} + s_{12}}h\right)$ is about $(-10...+10)$nm$^{-1}$. Since for the most of perovskites $Q_{12} < 0$, the conditions $\mu > 0$ and $u_m < 0$ favors the negative values of $\lambda_S^{(i)} \equiv \lambda_S$.

It is seen that renormalized inverse extrapolation length (14) linearly depends on both the intrinsic surface stress and misfit strain. This dependence is shown in Fig. 2 for different values of misfit strain. It is seen, that length $\lambda_S$ could be either negative or positive. Derived renormalization of the extrapolation length $\lambda$ given by (11) shows the possibility of negative length $\lambda_S$ appearance in the strained films even for the case when initially $\lambda > 0$. In particular the case $\lambda_S < 0$ can be realized at $\mu > 0$, $u_m \leq 0$ and $\lambda^{-1} < 0$, or $\lambda^{-1} > 0$, but the second term in Eq.(14) is usualy larger than the first. At $|\lambda| \leq 10^{-9}$m the length $\lambda_S$ diverges and changes its sign under the condition $(s_{11} + s_{12})\mu \approx 2u_m h$.

Thus, Eq.(11) provides the possible background of the negative extrapolation length appearance in strained perovskite thin films, previously postulated in the framework of Landau-Ginsburg phenomenology.

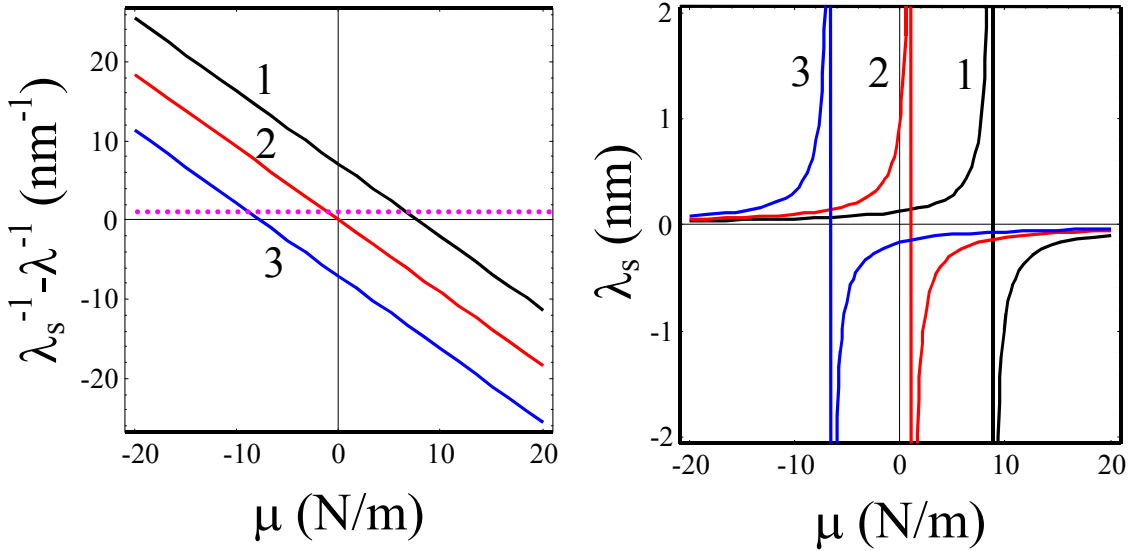

Fig. 2. Dependence of the inverse (a) and direct (b) extrapolation length on the intrinsic surface stress $\mu$, for misfit strain $u_m = 10^{-2}, 0, -10^{-2}$ (curves 1, 2, 3). Material parameters correspond to KTaO$_3$, namely $Q_{12} = -0.023\,\text{m}^4/\text{C}^2$, $s_{11} = 2.7 \cdot 10^{-12}\,Pa^{-1}$, $s_{12} = -0.62 \cdot 10^{-12}\,Pa^{-1}$. Other parameters: $g = 10^{-10}\,\text{V}\,\text{m}^3/\text{C}$, $h = 0.8 \cdot 10^{-9}\,\text{m}$, $\lambda = 10^{-9}\,\text{m}$.

The approximate solution of the nonlinear Euler-Lagrange equation for $P_3$ could be found by using the *direct variational method* as proposed earlier [20], [21], [22]. Using the solution of linearized Eq.(10): $P_3(z) = P_{3V}\dfrac{1-\varphi(z)}{1-\overline{\varphi(z)}} - \dfrac{P_{m1}+P_{m2}}{2}\varphi(z) - \dfrac{P_{m2}-P_{m1}}{2}\xi(z)$ (see Appendix B) as the trial function for the direct variational method, one can obtain free energy in conventional algebraic form with renormalized coefficients:

$$G \approx \alpha_T(T - T_S(l))\frac{P_{3V}^2}{2} + \left(a_{11} + \frac{Q_{12}^2}{s_{11}+s_{12}}\right)\frac{P_{3V}^4}{4} + a_{111}\frac{P_{3V}^6}{6} - P_{3V}(E_0 + E_m(l)) \quad (15)$$

The inhomogeneity (i.e. surface polarization $P_m$) in the boundary conditions (10) leads to the appearance of built-in field $E_m$. Under the typical conditions $\lambda_S\sqrt{\pi/g} \gg 1$ and $l\sqrt{\pi/g} \gg 1$ expression for $E_m$ has the form:

$$E_m(l) \approx \frac{4d_{31}^S u_m}{l(s_{11}+s_{12})}. \quad (16)$$

It is clear that built-in field $E_m \sim d_{31}^S u_m/l$ is inversely proportional to the film thickness $l$ and essentially increases with film thickness decrease [5]. More general case of three nonzero polarization components $P_{1,2,3}$ is considered in details in Ref.[22].

# 5. SIZE-INDUCED FERROELECTRICITY ENHANCEMENT IN CONVENTIONAL FERROELECTRIC FILMS

The exact expression for the ferroelectric phase transition temperature $T_S(l)$ in thin film on the substrate acquires the form:

$$T_S(l) = T_f(l, \lambda_S) + \frac{4Q_{12}u_m}{\alpha_T(s_{11} + s_{12})}. \tag{17}$$

Under the typical condition $l\sqrt{\pi/g} \gg 1$, Eq.(17) could be rewritten as: $T_S(l) \approx T_C^*(u_m) - \frac{2g}{\alpha_T l(\lambda_S + l_z/2)}$ where the temperature $T_C^*(u_m) \approx T_C^b + \frac{4Q_{12}u_m}{\alpha_T(s_{11} + s_{12})}$ and characteristic thickness $l_z = \sqrt{g/\pi}$ are introduced.

At a given temperature $T$ the film critical thickness $l_{cr}(T)$ (if any) should be found from the condition $T_S(l_{cr}) = T$. We obtained that $l_{cr}(T) \approx \frac{2g}{\alpha_T(T_C^* - T)(\lambda_S + l_z/2)}$; it exists at $-0.5l_z < \lambda_S < \infty$.

Size-induced transition temperature enhancement $T_S/T_C^* > 1$ is possible at negative lengths $-\infty < \lambda_S < -0.5l_z$. Usually $l_z \sim 5 - 0.5 \text{ A}^\circ$. The effect should be clearly distinguished from the thickness-independent renormalization $T_C^*(u_m)$ of $T_C^b$ obtained by Pertsev et al [3]. Size-induced transition temperature decrease $T_S/T_C^* < 1$ corresponds to the range of lengths $-0.5l_z < \lambda_S < \infty$. The value $\lambda_S \approx -0.5l_z$ is the special point. At positive $\lambda_S \geq l_z$ size effects are most pronounced in the region of thickness $l < 10^3 l_z$ that typically corresponds to the film of thickness less than 100nm. The relation between magnitudes and signs of intrinsic surface stress $\mu$ and misfit strain $u_m$, two quantities related to the confinement conditions, is clearly seen from Fig. 2.

It should be noted that only in the absence of electric field the temperature $T_S(l)$ determines the transition point, where all polar and dielectric properties have peculiarities. For instance, when the electric field is acting on the system, dielectric permittivity has maximum at temperature, different from transition temperature (17). Moreover, the built-in field $E_m$ (16), as well as external electric field, smears the sharp divergence of dielectric permittivity and shifts it to the higher temperatures, induces electret-like polar state at $l < l_{cr}$ (i.e. paraelectric phase is absent, see e.g. [5, 22]), especially in the case when $P_m$ is comparable with bulk spontaneous polarization $P_S$. In this case one can find temperature of dielectric permittivity maximum as follows:

$$T_m(l) = T_S(l) + \frac{3}{\alpha_T}\sqrt[3]{a_{11} + \frac{Q_{12}^2}{s_{11} + s_{12}}} \left(\frac{E_0 + E_m(l)}{4}\right)^{2/3} \tag{18}$$

Here we suppose that $a_{11} + Q_{12}^2/(s_{11} + s_{12}) > 0$ and neglect the contribution of $P_{3V}^6$ term from (15).

It is useful to introduce the characteristic length $l_m = 2d_{31}^S u_m/(E_c(s_{11} + s_{12}))$ that determines the relative magnitude of built-in field with respect to the thermodynamic coercive filed $E_c = 2(\alpha_T T_C^*(u_m))^{3/2}/\sqrt{27(a_{11} + Q_{12}^2/(s_{11} + s_{12}))}$ at $T=0$. It is clear that $E_m(l)/E_c \sim l_m/l$. The ratio $E_m(l)/E_c$ defines the ability of built-in field to be the source of the film self-polarization especially when it is more than unity.

Ferroelectric phase transition temperature $T_S/T_C^*$ and dielectric permittivity maximum temperature $T_m/T_C^*$ vs. dimensionless film thickness $l/l_z$ are depicted in Figs. 3 for different ratios $\lambda_S/l_z$. It is clear that always $T_m > T_S$ as it should be expected from Eq.(18), while the essential difference is related with the built-in field, the thinner the film the higher the difference. It is understood since the built-in filed is inversely proportional to the film thickness. Sometimes the maximum appeared on $T_m$ thickness dependence that is related with the competition of the depolarization field negative contribution $\sim (l_z/l)$ and positive built-in field contribution $\sim (l_m/l)^{2/3}$.

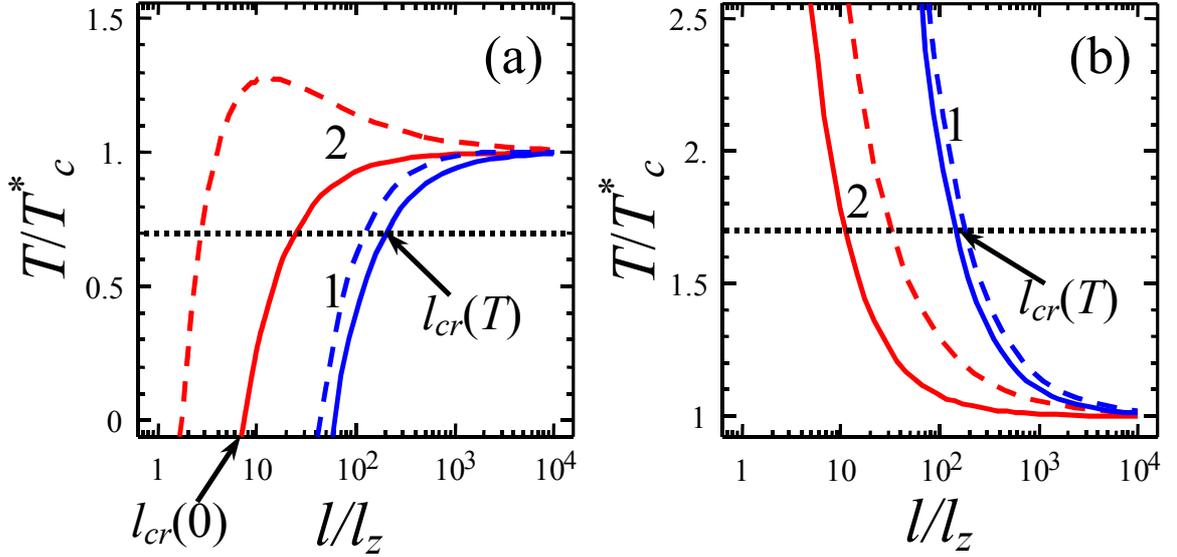

**FIG. 3.** Ferroelectric phase transition temperature $T_S/T_C^*$ (solid curves) and dielectric permittivity maximum temperature $T_m/T_C^*$ (dashed curves) vs. thickness $l/l_z$ for $l_m/l_z = 10$ and different values of extrapolation length, namely (a) $\lambda_S/l_z = 2, 20$ (curves 1 and 2) and (b) $\lambda_S/l_z = -2, -20$ (curves 1 and 2).

It is worth to underline that in the case $E_m \neq 0$ temperature (18) can be determined experimentally as the temperature of dielectric constant maximum, while temperature (17) can be found from independent experiments, e.g. it determines the point where the ferroelectric hysteresis loop vanishes.

## 6 SIZE-INDUCED FERROELECTRICITY IN INCIPIENT FERROELECTRIC FILMS

The Barrett formula $a_1(T) = \alpha_T \left( \frac{T_q}{2} \coth\left( \frac{T_q}{2T} \right) - T_0 \right)$ is valid for both *incipient* and *conventional* ferroelectrics [23] at wide temperature interval including low (quantum) temperatures. The transition temperature induced by surface and size effects is given by:

$$T_S(l) \approx \frac{T_q}{2} arc\coth^{-1}\left( \frac{2}{T_q}\left( T_0 + \frac{4Q_{12}u_m}{\alpha_T(s_{11}+s_{12})} - \frac{2g}{\alpha_T l(\lambda_S + l_z/2)} \right) \right) \quad (19)$$

At the temperatures $T \gg T_q/2$ one can obtain that $\alpha_T \left( \frac{T_q}{2} \coth\left( \frac{T_q}{2T} \right) - T_0 \right) \cong \alpha_T(T - T_0)$ and Eq.(19) in this limit gives Eq.(17) after substitution $T_C^b \to T_0$. Because of the fact we will consider $T_0$ as hypothetical phase transition temperature for incipient ferroelectrics.

It is worth to underline that the second term in brackets of Eq. (19) represents the contribution of stress originated from the mismatch effect that does not depend on the film thickness, while the third one is related to the depolarization field and correlation effect influence which define the thickness dependence of transition temperature.

Eq.(19) predicts the appearance of the *size-induced ferroelectric phase* thin films of incipient ferroelectrics at negative length $\lambda_S$ and/or compressive misfit strain $u_m \leq 0$ (since for many perovskites $Q_{12} < 0$). Ferroelectric phase transition temperature $T_S$ vs. film thickness $l$ calculated on the basis of Eq.(19) for KTaO$_3$ parameters is shown in Fig.4. Electrostriction constant $Q_{12}$ for KTaO$_3$ were taken from the work of Uwe and Sakudo [24] who carried out experiments on the influence of the external pressure on the dielectric properties of KTaO$_3$.

One can see from Fig. 4, that the thickness dependence of transition temperature, defined by the third term in Eq. (19), is essentially influenced by the value of $\lambda_S$. This term reflects the contribution of the surface and polarization gradient. The second term in Eq. (19), related to the influence of mismatch strain, lead to the shift of transition temperatures independent on the film thickness. For the case, presented in Fig. 4b, this shift is about 50 K to the higher temperatures (see the value at high $l$ or Eq. (19) at $l \to \infty$). Note, that this thickness independent shift was considered earlier for homogeneously strained thick films of SrTiO$_3$ taking into account three components of polarization and structural order parameter, but with size, surface and depolarization effects being ignored [4].

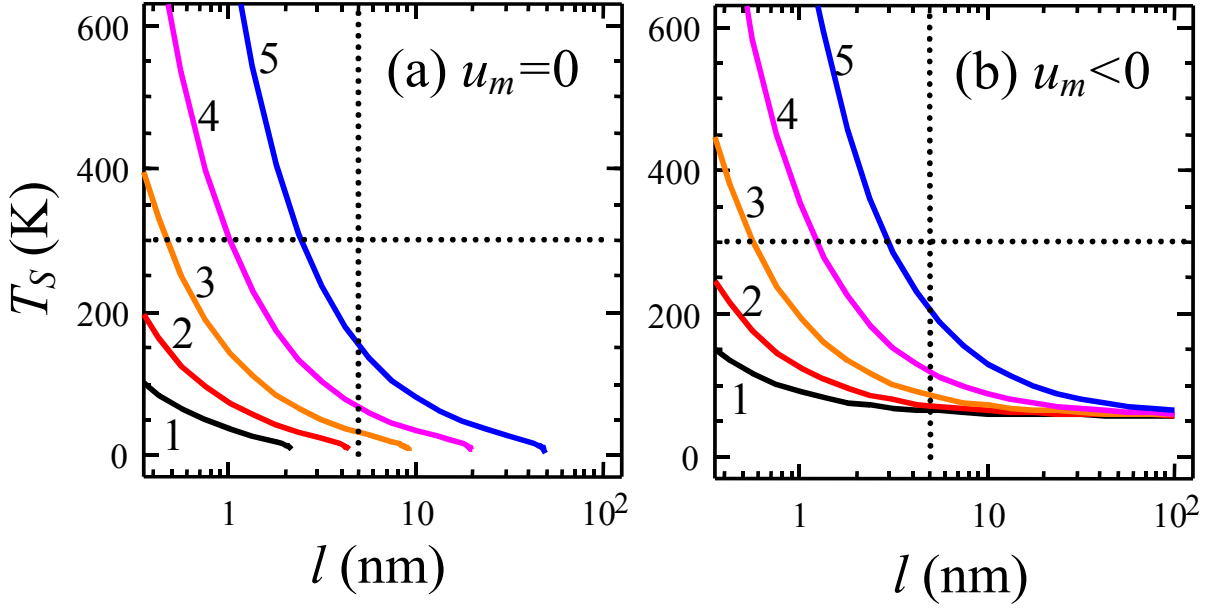

**FIG. 4.** Ferroelectric phase transition temperature $T_S$ vs. the film thickness $l$ for KTaO$_3$ material parameters $T_q = 55\,\text{K}$, $T_0 = 13\,\text{K}$, high temperature Curie-Weiss constant $C_{CW} = 5.6 \times 10^4\,\text{K}$, $Q_{12} = -0.023\,\text{m}^4/\text{C}^2$; gradient coefficient $g = 10^{-10}\,\text{V m}^3/\text{C}$ [19] in SI units, negative lengths $\lambda_S = -16;\,-8;\,-4;\,-2;\,-1$ Å (curves 1, 2, 3, 4, 5) aand **(a)** unstrained state ($u_m = 0$) and **(b)** compressed film ($u_m = -0.6\,\%$).

The temperature $T_S(l)$ determines the transition point in the absence of electric field. Similarly to the case of conventional ferroelectrics the built-in electric field $E_m$ (16) smears the sharp divergence of dielectric permittivity and shifts it to the higher temperatures, induces electret-like polar state at $l < l_{cr}$. In this case one can find temperature of dielectric permittivity maximum as follows:

$$T_m(l) \approx \frac{T_q}{2} \operatorname{arccoth}^{-1}\left( \frac{2}{T_q}\left( T_0 + \frac{4Q_{12}u_m}{\alpha_T(s_{11}+s_{12})} - \frac{2g}{\alpha_T l(\lambda_S + l_z/2)} + \frac{3}{\alpha_T}\sqrt[3]{a_{11} + \frac{Q_{12}^2}{s_{11}+s_{12}}}\left(\frac{E_m(l)}{4}\right)^{2/3} \right)\right) \quad (20)$$

The influence of built-in field is clear from Fig. 5. It is seen that the temperature of dielectric permittivity maximum is always higher than transition temperature, the thinner the film the higher the difference. It is understood since the built-in filed is inversely proportional to the film thickness. Moreover, for unstrained films there exist the region of thickness where the film is in paraelectric phase (i.e. the ferroelectric hysteresis loop is absent), while the permittivity reveals the temperature maximum.

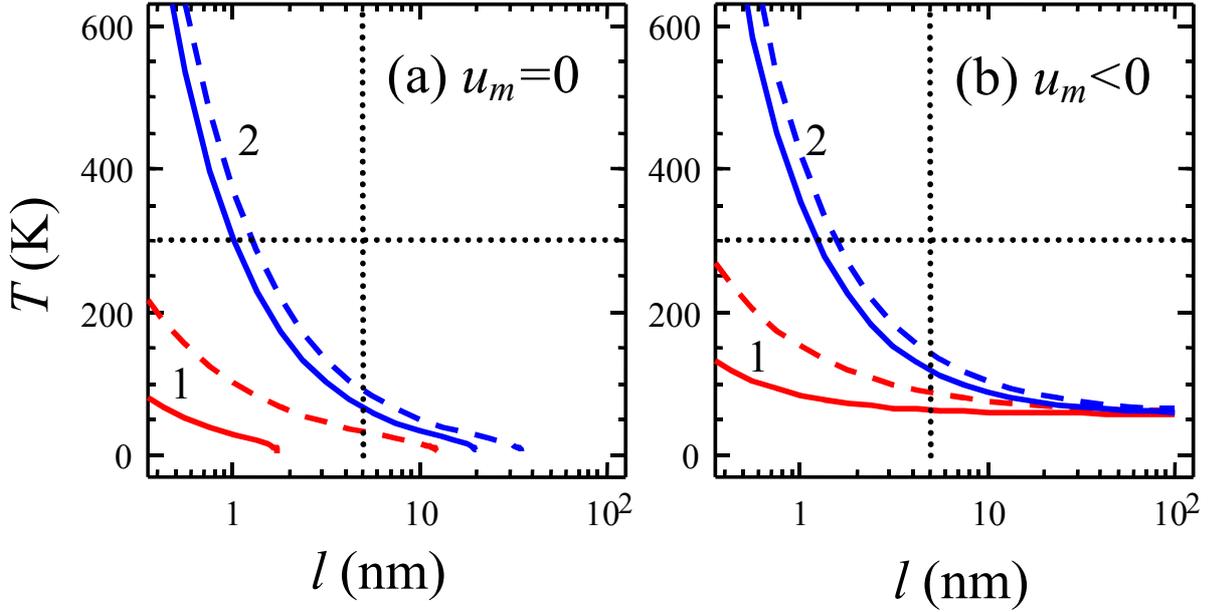

**FIG. 5.** Ferroelectric phase transition temperature $T_S/T_C^*$ (solid curves) and temperature of dielectric permittivity maximum $T_m/T_C^*$ (dashed curves) vs. thickness $l/l_z$ for (a) unstrained films ($u_m = 0$) and (b) compressed film ($u_m = -0.6\%$), with KTaO$_3$ material parameters (see caption to Fig. 4), $l_m = 4\,nm$ and the following values of extrapolation length $\lambda_S/l_z = -20, -2$ Å (curves 1 and 2).

## CONCLUSION

In our consideration we came to the conclusion about size-driven ferroelectricity in the incipient ferroelectrics KTaO$_3$ thin films of thickness less then 50 nm at room temperature.

The effects are expected to be the most pronounced for compressed films ($u_m \leq 0$) and negative lengths $\lambda_S$. The obtained expression for renormalized extrapolation length $\lambda_S$ allows one to explain the physical reasons for negative or positive $\lambda_S$ values, which were simply postulated in the majority of earlier papers. In particular the condition $\left(\mu^S Q_{12} - \dfrac{2q_{12}^S u_m}{s_{11}+s_{12}} + \dfrac{a_1^S}{4}\right) < 0$ on the components of electrostriction tensor $Q_{ij}$, surface stress tensor $\mu^S$, the surface electrostriction tensor $q_{12}^S$, the elastic stiffness tensor $s_{ij}$, misfit strain $u_m$ and surface dielectric stiffness $a_1^S$ was shown to be necessary for negative sign of $\lambda_S$. This means that polarization at the surface is larger than that in the bulk.

Resuming, it is worth to underline that the simultaneous consideration of intrinsic surface stress, misfit strain (if any) polarization gradient and depolarization field are rather important for the adequate description and prediction of size-induced ferroelectricity enhancement in conventional thin ferroelectric films and its appearance in incipient nanosized ferroelectrics. It is obvious that the similar surface effects can influence strongly on the phase transitions in the thin films of another incipient ferroelectrics $SrTiO_3$. The theoretical prediction could be important for the modern applications.

**Appendix A. Free energy of thin unipolar perovskite film on a rigid substrate**

Let us consider ferroelectric thin film with the thickness $l$ $(-l/2 \le z \le l/2)$ on the thick substrate in the external electric field $\mathbf{E}$. Helmholtz free energy expansion on polarization $\mathbf{P} = (0, 0, P_3)$, stress $\sigma_{nm}$ and strain $u_{jk}$ powers has the form:

$$F_V = S \int_{-l/2}^{l/2} dz \left( \begin{array}{l} \dfrac{a_1}{2} P_3^2 + \dfrac{a_{11}}{4} P_3^4 + \dfrac{a_{111}}{6} P_3^6 + \dfrac{g}{2}\left(\dfrac{dP_3}{dz}\right)^2 - P_3\left(E_0 + \dfrac{E_3^d}{2}\right) - \\ -(Q_{12}(\sigma_{11}+\sigma_{22}) + Q_{11}\sigma_{33})P_3^2 - \dfrac{1}{2}s_{11}(\sigma_{11}^2+\sigma_{22}^2+\sigma_{33}^2) - \\ -s_{12}(\sigma_{11}\sigma_{22}+\sigma_{11}\sigma_{33}+\sigma_{33}\sigma_{22}) - \dfrac{1}{2}s_{44}(\sigma_{32}^2+\sigma_{31}^2+\sigma_{12}^2) + u_{jk}\sigma_{jk} \end{array} \right) \quad (A.1)$$

The surface free energy expansion (1.1b) acquires the form

$$F_S = \sum_{i=1}^{2} \int_{S_i} dxdy \left( \begin{array}{l} \dfrac{a_1^i}{2} P_3^2 + \dfrac{a_{11}^i}{4} P_3^4 + \tau^i(\sigma_{11}+\sigma_{22}+\sigma_{33}) + \dfrac{\eta_1}{2}(\sigma_{11}+\sigma_{22}+\sigma_{33})^2 + \\ + \dfrac{\eta_2}{2}(\sigma_{11}^2+\sigma_{22}^2+\sigma_{33}^2 + 2\sigma_{12}^2 + 2\sigma_{32}^2 + 2\sigma_{13}^2) + \\ + d_{31}^i(\sigma_{11}+\sigma_{22})P_3 + d_{33}^i\sigma_{33}P_3 - q_{12}^{Si}(\sigma_{11}+\sigma_{22})P_3^2 - q_{11}^i\sigma_{33}P_3^2 \end{array} \right). \quad (A.2)$$

Where the perovskite symmetry ($Q_{1133} = Q_{2233} = Q_{12}$, $Q_{3333} = Q_{33} = Q_{11}$, $d_{311}^i = d_{322}^i = d_{31}^i$, $d_{333}^i = d_{33}^i$) and Voigt notation or matrix notation (xx=1, yy=2, zz=3, zy=4, zx=5, xy=6) are used.

We assume that at the film-substrate surface ($z = -l/2$) the strain is induced by film-substrate lattice mismatch as following:

$$u_{11} = u_{22} = u_m, \quad u_{12} = 0. \quad (A.3)$$

At the film-ambient free surface ($z = +l/2$) the stress is determined as

$$\sigma_{33} = \sigma_{31} = \sigma_{32} = 0. \quad (A.4)$$

In the considered case of a single-domain perovskite film on a rigid cubic substrate, internal elastic fields are homogeneous so that the above conditions hold throughout the film volume. Thus, the solution of the "mixed" elastic problem, obtained after minimization of $\dfrac{\partial G_V}{\partial \sigma_{jk}} = -u_{jk}$ has the form:

$$u_{11} = u_{22} = u_m, \quad u_{33} = 2s_{12}\dfrac{u_m - Q_{12}P_3^2}{s_{11}+s_{12}} + Q_{11}P_3^2, \quad u_{23} = u_{13} = u_{12} = 0, \quad (A.5a)$$

$$\sigma_{11} = \sigma_{22} = \dfrac{u_m - Q_{12}P_3^2}{s_{11}+s_{12}}, \quad \sigma_{12} = \sigma_{31} = \sigma_{32} = \sigma_{33} = 0. \quad (A.5b)$$

## Appendix B. Trial function for thin strained films

Under the condition $\lambda_{S1} = \lambda_{S2} = \lambda_S$ linearized solution have the form [22]:

$$P_3(z) = \frac{E_0 - 2\pi\overline{\varphi(z)}P_m}{\alpha_T(T - T_C^*) + 4\pi\overline{\varphi(z)}}[1 - \varphi(z)] - \frac{P_m}{2}[\varphi(z) - \xi(z)] \quad (B.1)$$

Here $\varphi(z) = \dfrac{ch(z/l_z)}{ch(l/2l_z) + (\lambda_S/l_z)sh(l/2l_z)}$, $\xi(z) = \dfrac{sh(z/l_z)}{sh(l/2l_z) + (\lambda_S/l_z)ch(l/2l_z)}$ and $l_z \approx \sqrt{g/4\pi}$.

Averaged susceptibility $\overline{\chi_{33}(z)} = \dfrac{1 - \overline{\varphi(z)}}{\alpha_T(T - T_C^*) + 4\pi\overline{\varphi(z)}}$, thus renormalized coefficient

$A_3 \equiv \dfrac{1}{\overline{\chi_{33}(z)}} = \left(\dfrac{\alpha_T(T - T_C^*) + 4\pi\overline{\varphi(z)}}{1 - \overline{\varphi(z)}}\right)$. It is easy to obtain that:

$$\overline{\varphi}(l) = \frac{2l_z}{l} \cdot \frac{sh(l/2l_z)}{ch(l/2l_z) + (\lambda_S/l_z)sh(l/2l_z)} \approx \begin{cases} \dfrac{1}{h(1+\Lambda_S)}, & h \gg 1 \\ \dfrac{1}{1+\Lambda_S h}, & h \ll 1 \end{cases} \quad (B.2)$$

Hereinafter we used designations $h = \dfrac{l}{2l_z}$, $\Lambda_S = \dfrac{\lambda_S}{l_z}$ and variational function:

$$P_3(z) = P_V \frac{1 - \varphi(z)}{1 - \overline{\varphi(z)}} - \frac{P_m}{2}[\varphi(z) - \xi(z)].$$

**References**

1. D. R. Tilley, *Finite size effects on phase transitions in ferroelectrics*. in: *Ferroelectric Thin Films*, ed. C. Paz de Araujo, J. F.Scott, and G. W. Teylor (Gordon and Breach, Amsterdam, 1996) 11.

2 O. Auciello, Science and technology of thin films and interfacial layers in ferroelectric and high-dielectric constant heterostructures and application to devices. J. Appl. Phys. **100**, 051614 (2006).

3 N. A. Pertsev, A. G. Zembilgotov, and A. K. Tagantsev, Effect of Mechanical Boundary Conditions on Phase Diagrams of Epitaxial Ferroelectric Thin Films. *Phys. Rev. Lett*. **80**, 1988 (1998).

4 N. A. Pertsev, A. K. Tagantsev, and N. Setter. Phase transitions and strain-induced ferroelectricity in SrTiO$_3$ epitaxial thin films. Phys. Rev. B **61**, №2, R825 - 829 (2000).

5 M. D. Glinchuk and A. N. Morozovska, The internal electric field originating from the mismatch effect and its influence on ferroelectric thin film properties. J. Phys.: Condens. Matter **16**, 3517 (2004).

6 Shchukin V.A., Bimberg D., Spontaneous ordering of nanostructures on crystal surfaces. Rev. Mod. Phys. **71**(4), 1125-1171 (1999).

7 J. H. Haeni, P. Irvin, W. Chang, R. Uecker, P. Reiche, Y. L. Li, S.BChoudhury, W. Tian, M. E. Hawley, B. Craigo, A. K. Tagantsev, X. Q. Pan, S. K. Streiffer, L. Q. Chen, S. W. Kirchoefer, J. Levy, and D. G. Schlom, Nature (London) **430**, 758 (2004).

8 Y. L. Li, S. Choudhury, J. H. Haeni, M. D. Biegalski, A. Vasudevarao, A. Sharan, H. Z. Ma, J. Levy, Venkatraman Gopalan, S. Trolier-McKinstry, D. G. Schlom, Q. X. Jia, and L. Q. Chen, Phase transitions and domain structures in strained pseudocubic (100) $SrTiO_3$ thin films, Phys. Rev. B 73, 184112 (2006).

9 L.D. Landau and E.M. Lifshitz, *Statistical physics. Theoretical Physics*, Vol. 5 (Butterworth-Heinemann, Oxford, 1998)

10 L.D. Landau and E.M. Lifshitz, *Electrodynamics of Continuous Media*, (Butterworth Heinemann, Oxford, 1980).

11 I.V.Marchenko, and A.Ya.Parshin, About elastic properties of the surface of crystals. Zh. Eksp. Teor. Fiz. **79** (1), 257-260 (1980), [Sov. Phys. JETP **52**, 129-132 (1980)].

12 A.K. Tagantsev, Piezoelectricity and flexoelectricity in crystalline dielectrics. Phys. Rev. B, **34**, 5883 (1986).

13 A. M. Bratkovsky, and A. P. Levanyuk. Smearing of Phase Transition due to a Surface Effect or a Bulk Inhomogeneity in Ferroelectric Nanostructures. *Phys. Rev. Lett*. **94**, 107601 (2005).

14 L.D. Landau and E.M. Lifshitz, Theory of Elasticity. Theoretical Physics, Vol. 7 (Butterworth-Heinemann, Oxford, U.K., 1998)

15 J. S. Speck, W. Pompe, Domain configurations due to multiple misfit relaxation mechanisms in epitaxial ferroelectric thin films. I. Theory. J. Appl. Phys. **76**, 466 (1994).

16 M.Saad, P.Baxter, R.M.Bowman, J.M.Gregg, F.D.Morrison, and J.F. Scott Intrinsic dielectric response in ferroelectric nano-capacitors. *J. Phys.: Condens. Matter* **16,** L451-454 (2004).

17 R. Kretschmer and K. Binder, Surface effects on phase transition in ferroelectrics and dipolar magnets. *Phys. Rev*. B **20**, 1065 (1979).

18 W. Ma, M. Zhang, and Z. Lu, A study of size effects in $PbTiO_3$ nanocrystals by Raman spectroscopy. Phys. Stat. Sol. (a) **166**, 811-815 (1998).

19 G. B. Stephenson, and K. R. Elder. Theory for equilibrium 180° stripe domains in $PbTiO_3$ films. J. Appl. Phys. **100**, 051601 (2006)

20 M.D.Glinchuk, E.A.Eliseev, and V.A.Stephanovich, The depolarization field effect on the thin ferroelectric films properties. Physica B, **332**, 356 (2002).

21 M.D.Glinchuk, E.A.Eliseev, V.A.Stephanovich, and R. Fahri. Ferroelectric thin films properties - Depolarization field and renormalization of a "bulk" free energy coefficients. J. Appl. Phys. **93**, 1150 (2003).

22 M. D.Glinchuk, A. N. Morozovska, E. A. Eliseev, Ferroelectric thin films phase diagrams with self-polarized phase and electret state. J. Appl. Phys. – 2006. – Vol. **99**, № 11, pp. 114102-1-12.

23 J. H. Barrett, Dielectric Constant in Perovskite Type Crystals. Phys. Rev. **86**, 118 (1952).

24 H. Uwe and T. Sakudo, Raman-scattering study of stress-induced ferroelectricity in $KTaO_3$. Phys. Rev. B **15**, №1, 337 - 345 (1977).